\documentclass[conference]{IEEEtran}

\usepackage{graphicx}
\usepackage{color}
\usepackage{placeins}
\usepackage{float}
\usepackage{tabularx,colortbl}
\usepackage{epstopdf}
\usepackage{amsmath}
\usepackage{amssymb}
\usepackage{caption}
\usepackage{subcaption}
\usepackage{algorithm}
\usepackage{algorithmic}
\usepackage{multirow}
\usepackage{url}

\renewcommand{\baselinestretch}{.961}

\begin{document}
%
\title{Power Control and Mode Selection for VBR Video Streaming in D2D Networks}
%
%
\author{\authorblockN{Chuang Ye, M. Cenk Gursoy, and Senem Velipasalar}\\
\authorblockA{Department of Electrical Engineering and Computer Science\\
Syracuse University, Syracuse, NY 13244\\ Email: chye@syr.edu, mcgursoy@syr.edu, svelipas@syr.edu}}

\maketitle

\begin{abstract}
\let\thefootnote\relax\footnote{This work was supported in part by National Science Foundation grant CCF-1618615, ECCS-1443994 and CNS-1443966.}
In this paper, we investigate the problem of power control for streaming variable-bit-rate (VBR) videos in a device-to-device (D2D) wireless network. A VBR video traffic model that considers video frame sizes and playout buffers at the mobile users is adopted. A setup with one pair of D2D users (DUs) and one cellular user (CU) is considered and three modes, namely cellular mode, dedicated mode and reuse mode, are employed. Mode selection for the data delivery is determined and the transmit powers of the base station (BS) and device transmitter are optimized with the goal of maximizing the overall transmission rate while VBR video data can be delivered to the CU and DU without causing playout buffer underflows or overflows. A low-complexity algorithm is proposed. Through simulations with VBR video traces over fading channels, we demonstrate that video delivery with mode selection and power control achieves a better performance than just using a single mode throughout the transmission.
\end{abstract}
\thispagestyle{empty}

\begin{IEEEkeywords} device-to-device communication, variable bit rate video traffic, mode selection, power control. \end{IEEEkeywords}

%
\IEEEpeerreviewmaketitle

\section{Introduction}
Recently, multimedia applications such as video telephony, teleconferencing, and video streaming have started becoming predominant in data transmission over wireless networks. For instance, as reported in \cite{Cisco}, mobile video traffic exceeded 50\% of the total mobile data traffic for the first time in 2012, and grew to 60\% in 2016, and more than three-fourths of the global mobile data traffic is expected to be video traffic by 2021. In such multimedia applications, certain quality of service (QoS) guarantees need to be provided in order to satisfy the performance requirements of the end-users, and this, in general, has to be accomplished with only limited wireless resources.

Recently, scheduling algorithms to transmit multiple video streams from a base station to mobile clients were investigated in \cite{seetharma}. With the proposed algorithms, the vulnerability to stalling was reduced by allocating slots to videos in a way that maximizes the minimum “playout lead” across all videos with an epoch-by-epoch framework. The distribution of prefetching delay and the probability generating function of playout buffer starvation for constant bit rate (CBR) streaming was modeled in \cite{xu}. The framework to characterize the throughput variation caused by opportunistic scheduling at the BS, and the playback variation of variable bit rate (VBR) traffic were considered as an extension. The flow dynamics have dominant influence on QoE metrics compared to the variation of throughput caused by fast channel fading and that of video playback rate caused by VBR streaming. Authors in \cite{yhuang} proposed algorithms to find the optimal transmit powers for the base stations with the goal of maximizing the sum transmission rate while VBR video data can be delivered to mobile users without causing playout buffer underflows or overflows. A deterministic model for
VBR video traffic that considers video frame sizes and playout buffers at the mobile users was adopted. \cite{chatziperis}, and reference \cite{rango} investigated effective admission control schemes for VBR videos over wireless networks in terms of bandwidth and QoS requirements.

In this paper, we consider the problem of streaming VBR videos in D2D wireless networks. VBR video has stable video quality across frames at the cost of large variations in the frame size or bit rate, where CBR video has a stable bit rate but the visual qualities of the frames may vary significantly. We consider a deterministic traffic model for stored VBR video, taking into account the frame size, frame rate, and playout buffers as in \cite{sen} and \cite{liang}. We exploit power control and the transmission mode selection with the goal of maximizing the sum transmission rate while avoiding underflows and overflows under transmit power constraints in a D2D wireless network.

The remainder of this paper is organized as follows: The system model is described in Section \ref{sec:System_Model}. Optimization problems are formulated and the optimal policies are derived in Sections \ref{sec:Formation} and \ref{sec:Strategy}. Simulation results are presented and discussed in Section \ref{sec:Result}. Finally, we conclude the paper in Section \ref{sec:Conclusion}.

\section{System Model} \label{sec:System_Model}
\begin{figure}
\centering
\includegraphics[width=0.3\textwidth]{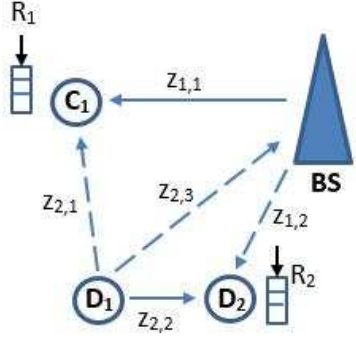}
\caption{\small{Proposed system block diagram for VBR video streaming in D2D networks}}\label{fig:System_Model}
\end{figure}

As mentioned above, with the goal of maximizing the overall transmission rate at the device user (DU) and cellular user (CU), we study optimal strategies for mode selection and resource allocation in a cellular network with D2D pairs operating under peak transmission power and buffer underflow and overflow constraints. For simplicity, we consider a D2D cellular wireless transmission network with a single base station (BS), which serves one CU denoted by $\{C_1\}$ as illustrated in Fig. \ref{fig:System_Model}. There also exists a pair of DUs denoted by $\{D_1, D_2\}$. We assume that the transmissions between D2D users and also between cellular user and BS are one-way, i.e., $BS$ and $D_1$ are transmitters while $C_1$ and $D_2$ are the receivers. The maximum transmit powers of the two transmitters, namely the $BS$ and $D_1$, are denoted by $P_{bmax}$ and $P_{dmax}$, respectively. In the cellular link, $BS$ sends information to $C_1$ via the downlink channel. In the D2D link, $D_1$ transmits data to $D_2$ either directly or via the $BS$ depending on the mode selection. The data packets are stored in buffers at the receivers before playout. Underflow and overflow constraints are imposed on these receiving buffers. The total bandwidth is denoted as $B$, and three modes, namely cellular mode, dedicated mode and reuse mode are employed in the system. The bandwidth is equally allocated, i.e., the bandwidth allocated to each link is denoted as $B_c = \frac{B}{3}$, $B_d = \frac{B}{2}$ and $B_r = B$ in cellular mode, dedicated mode and reuse mode, respectively.

\begin{figure}
\centering
\includegraphics[width=0.4\textwidth]{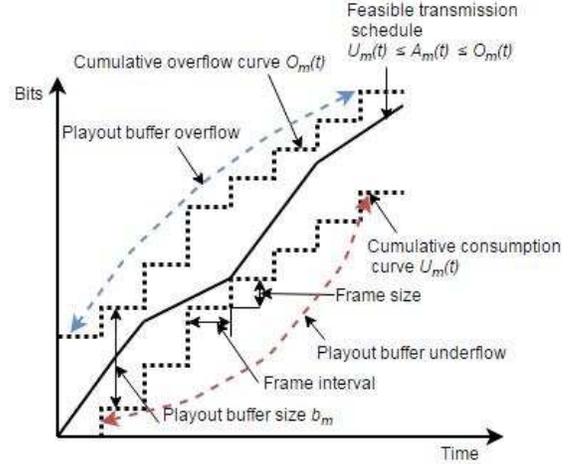}
\caption{\small{Cumulative overflow and cumulative consumption curves and a feasible transmission schedule for video}}\label{fig:Buffer}
\end{figure}

Let $z_{1,1}(t)$, $z_{1,2}(t)$, $z_{2,1}(t)$, $z_{2,2}(t)$ and $z_{2,3}(t)$ denote the instantaneous channel power gains of the links $BS$-$C_1$, $BS$-$D_2$, $D_1$-$C_1$, $D_1$-$D_2$ and $D_1$-$BS$ at time $t$, respectively. And also let $U_1(t)$ and $U_2(t)$ be the cumulative data consumption curves at the receiving users $C_1$ and $D_2$, representing the cumulative amount of data consumed by the decoders at time $t$, respectively. The cumulative data consumption curve is determined by the video characteristics such as frame sizes and rates, and the playout schedule. It is assumed that the playout buffers of $C_1$ and $D_2$ have sizes of $b_1$ and $b_2$ bits, and their videos have $L_1$ and $L_2$ frames, respectively. The cumulative overflow curve is formulated as \cite{yhuang}
\begin{equation}
O_m(t) = \min\{U_m(t-1)+b_m, U_m(L_m)\}, 0\leq t \leq L_m ,
\end{equation}
where $O_m(t)$ is the maximum amount of accumulated received bits at time $t$ without an overflow in the playout buffer. The cumulative transmission curves $A_1(t)$ and $A_2(t)$ are defined as the cumulative amount of bits received at $C_1$ and $D_2$ at time slot $t$, respectively. For simplicity, it is assumed that the videos have identical frame rates and the frame intervals are synchronized, which means that a time slot $t$ is the same as the $t$-th frame interval, for $0 \leq t \leq \max_m\{L_m\}$.
Since $O_m(t)$, $U_m(t)$ and $A_m(t)$ are cumulative curves, they are all nondecreasing functions over time. Fig. \ref{fig:Buffer} shows that the feasible transmission schedule needs to generate a cumulative transmission curve $A_m(t)$ that lies within $O_m(t)$ and $U_m(t)$ in order to play the video without stall events or overflows leading to missing frames.

\section{Problem Formation}\label{sec:Formation}
We consider a block fading channel with channel gains not changing within each time slot, but varying over different time slots following a certain distribution. Without loss of generality, we employ the Shannon capacity as the transmission rate in the D2D wireless network. Three different transmission modes are considered next.

\subsection{Cellular Mode}
In the cellular mode, $D_1$ sends the video data to $D_2$ via $BS$, which is acting as a relay node. Note that $BS$ also sends data to $C_1$ directly without any interference.  Since we need to guarantee that all the data transmitted from $D_1$ is received at $D_2$, $BS$ needs to deliver all the received bits from $D_1$ to $D_2$. Therefore, the transmission rate $R_1(t)$ from $BS$ to $C_1$ and the rate $R_2(t)$ from $D_1$ to $D_2$ via $BS$ over a two-hop link are derived as follows:
\begin{align}
& R_{1,1}(t, \mathbf{P}) = B_c \log \left(1+\frac{P_{b1}(t) z_{1,1}(t)}{N_0B_c}  \right) \label{R_c1}\\
& R_{2,1}(t, \mathbf{P}) = \min \{R_{3,1}(t), R_{4,1}(t)\} \label{R_c2}
\end{align}
where
\begin{align}
& R_{3,1}(t, \mathbf{P}) = B_c \log \left(1+\frac{P_d(t) z_{2,3}(t)}{N_0B_c} \right) \label{R_c3} \\
& R_{4,1}(t, \mathbf{P}) = B_c \log \left(1+\frac{P_{b2}(t) z_{1,2}(t)}{N_0B_c}  \right) \label{R_c4}
\end {align}
are the transmission rates from $D_1$ to $BS$ and from $BS$ to $D_2$, respectively. $\mathbf{P}$ is the transmit power vector $[P_{b1}(t), P_{b2}(t), P_d(t)]$.

\subsection{Dedicated Mode}
In dedicated mode, $D_1$ transmits data to $D_2$ directly over a separate channel without any interference. The transmission rates of $BS$-$C_1$ and $D_1$-$D_2$ links are given, respectively, by
\begin{align}
& R_{1,2}(t, \mathbf{P}) = B_d \log \left(1+\frac{P_{b1}(t) z_{1,1}(t)}{N_0B_d}  \right) \label{R_d1}\\
& R_{2,2}(t, \mathbf{P}) = B_d \log \left(1+\frac{P_{d}(t) z_{2,2}(t)}{N_0B_d}  \right). \label{R_d2}
\end{align}

\subsection{Reuse Mode}
In reuse mode, $D_1$ transmits data to $D_2$ directly but this time interference is experienced since $BS$-$C_1$ and $D_1$-$D_2$ links use the same channel. The transmission rates of $BS$-$C_1$ and $D_1$-$D_2$ links are
\begin{align}
& R_{1,3}(t, \mathbf{P}) = B_r \log \left(1+\frac{P_{b1}(t) z_{1,1}(t)}{P_{d}(t)z_{2,1}(t)+N_0B_r}  \right) \label{R_r1}\\
& R_{2,3}(t, \mathbf{P}) = B_r \log \left(1+\frac{P_{d}(t) z_{2,2}(t)}{P_{b1}(t)z_{1,2}(t)+N_0B_r}  \right). \label{R_r2}
\end{align}

Once the arrival rate is determined, $R_{m,n}(t, \mathbf{P})\tau$ video bits will be transmitted to the corresponding receiver in that time slot, where the subscript $n \in \mathcal{N} = {1, 2, 3}$ indicates the cellular mode, dedicated mode and reuse mode, respectively. The cumulative transmission curve $A_m(t)$ can be written as
\begin{align}
A_m(0) = 0, A_m(t) = A_m(t-1) + R_{m, n}(t, \mathbf{P})\tau. \label{A_1}
\end{align}

It is assumed that the peak power is $P_{dmax}$ at $D_1$, and $P_{bmax}$ at $BS$ for the cellular link $BS$-$C_1$ and downlink $BS$-$D_2$. The problem is to determine the transmit power vector $\mathbf{P}$ and to select the transmission mode for $0 < t \leq \max_m\{L_m\}$, such that the resulting cumulative transmission curves satisfy
\begin{align}
U_m(t) \leq A_m(t) \leq O_m(t), \forall m, t, \label{A_2}
\end{align}
i.e., no playout buffer underflow or overflow occurs at $C_1$ and $D_2$.
From (\ref{A_1}) and (\ref{A_2}), the feasible transmission rate range is
\begin{align}
\max\{0, \alpha_m(t)\} \leq R_{m,n}(t, \mathbf{P}) \leq \beta_m(t), \label{R_lim}
\end{align}
where $\alpha_m(t) = \frac{U_m(t)-A_m(t-1)}{\tau}$ and $\beta_m(t) = \frac{O_m(t) - A_m(t-1)}{\tau}$.

Let $R_{tot, n}(t, \mathbf{P}) = R_{1, n}(t, \mathbf{P}) + R_{2, n}(t, \mathbf{P})$ be the total transmission rate in 3 different modes. The optimal power control and mode selection problem for VBR video streaming is formulated as follows:
\begin{align}
\max_{n\in \mathcal{N}, \mathbf{P}} & R_{tot}(t, \mathbf{P}) = \sum_{m=1}^{2}R_{m, n}(t, \mathbf{P}) \\
s.t. \quad & P_{b1}(t) < P_{bmax} \label{P_b1}\\
& P_{b2}(t) < P_{bmax} \label{P_b2} \\
& P_d(t) < P_{dmax} \label{P_d} \\
& R_{m,n}(t, \mathbf{P}) \geq \max\{0, \alpha_m(t)\}, \forall m, \label{A_4}\\
& R_{m,n}(t, \mathbf{P}) \leq \beta_m(t), \forall m. \label{A_3}
\end{align}

\section{Optimal Power Control and Mode Selection Strategies}\label{sec:Strategy}
In this section, the optimal power control strategies in 3 different modes are identified and the best one among these 3 strategies is chosen as the final decision.
\subsection{Cellular Mode}
In cellular mode, there is no interference among the cellular link $BS$-$C_1$, uplink $D_1$-$BS$ and downlink $BS$-$D_2$ since these 3 links are operating in different channels. Hence, the maximum sum rate of $R_{tot, 1}(t)$ is the sum of maximum rates $R_{1,1}(t)$ and $R_{2,1}(t)$. From (\ref{R_c1}) and (\ref{A_3}), and the maximum power constraint (\ref{P_b1}), the maximum transmission rate through link $BS$-$C_1$ is determined as
\begin{align}
R_{c1}^*(t, \mathbf{P}) = \min \bigg\{B_c \log \left(1+\frac{P_{bmax} z_{1,1}(t)}{N_0B_c}  \right), \beta_1(t)\bigg\}.
\end{align}
Then, the optimal transmission power at $BS$ for the cellular link is found as
\begin{align}
P_{b1,1}^* = \left(2^{\frac{R_{c1}^*(t, \mathbf{P})}{B_c}}-1\right)\frac{N_0 B_c}{z_{1,1}(t)}.  \label{P_b1_c}
\end{align}
Considering (\ref{R_c2}) and (\ref{A_3}), and the maximum power constraints (\ref{P_b2}) and (\ref{P_d}), we determine the maximum transmission rate through the two-hop link $D_1$-$BS$-$D_2$ as
\begin{align}
R_{c2}^*(t, \mathbf{P}) = & \min \bigg\{B_c \log \left(1+\frac{P_{dmax} z_{2,3}(t)}{N_0B_c} \right), \nonumber \\
 & B_c \log \left(1+\frac{P_{bmax} z_{1,2}(t)}{N_0B_c}  \right), \beta_2(t)\bigg\}.
\end{align}
Hence, the optimal transmission powers in the uplink and downlink are derived as follows:
\begin{align}
P_{b2,1}^* & = \left( 2^{\frac{R_{c2}^*(t, \mathbf{P})}{B_c}}-1 \right) \frac{N_0 B_c}{z_{1,2}(t)}  \label{P_b2_c}      \\
P_{d,1}^* & = \frac{z_{1,2}(t)}{z_{2,3}(t)} P_{b2,1}^*.  \label{P_d_c}
\end{align}
Thus, the optimal transmit power vector is $\mathbf{P}_1 = [P_{b1, 1}^*, P_{b2,1}^*, P_{d,1}^*]$.

There are three scenarios based on the values of $R_{c1}^*(t, \mathbf{P})$ and $R_{c2}^*(t, \mathbf{P})$.
\begin{enumerate}

\item First, let us assume that $R_{c1}^*(t, \mathbf{P}) \geq \alpha_1(t)$ and $R_{c2}^*(t, \mathbf{P}) \geq \alpha_2(t)$, which means that the underflow and overflow playout buffer constraints at both $C_1$ and $D_2$ are satisfied. This has the highest priority, and let $pri(1) = 1$;

\item Second case is that either $R_{c1}^*(t, \mathbf{P}) \geq \alpha_1(t)$ or $R_{c2}^*(t, \mathbf{P}) \geq \alpha_2(t)$, which means that the underflow and overflow playout buffer constraints at either $C_1$ or $D_2$ are satisfied only. The priority of this scenario is lower, and let $pri(1) = 2$;

\item The last scenario is that neither conditions ($R_{c1}^*(t, \mathbf{P}) \geq \alpha_1(t)$ and $R_{c2}^*(t, \mathbf{P}) \geq \alpha_2(t)$) are satisfied, which means that the underflow and overflow playout buffer constraints at both $C_1$ and $D_2$ are not satisfied. This case has the lowest priority, and let $pri(1) = 3$;

\end{enumerate}

\subsection{Dedicated Mode}
Similarly as in cellular mode, from (\ref{R_d1}), (\ref{A_3}) and the constraint in (\ref{P_b1}), the maximum transmission rate in link $BS$-$C_1$ is
\begin{align}
R_{d1}^*(t, \mathbf{P}) = \min \bigg\{B_d \log \left(1+\frac{P_{bmax} z_{1,1}(t)}{N_0B_d}  \right), \beta_1(t)\bigg\},
\end{align}
and the optimal transmission power at $BS$ in cellular link is
\begin{align}
P_{b1,2}^* = \left(2^{\frac{R_{d1}^*(t)}{B_d}}-1\right)\frac{N_0 B_d}{z_{1,1}(t)}.  \label{P_b1_d}
\end{align}
Since $D_1$ transmits data to $D_2$ directly in the dedicated mode, the maximum transmission rate in direct link $D_1$-$D_2$ is
\begin{align}
R_{d2}^*(t, \mathbf{P}) = \min \bigg\{B_d \log \left(1+\frac{P_{dmax} z_{2,2}(t)}{N_0B_d}  \right), \beta_2(t) \bigg\},
\end{align}
and the optimal transmission power at $BS$ in cellular link is
\begin{align}
P_{d,2}^* = \big(2^{\frac{R_{d2}^*(t)}{B_d}}-1\big)\frac{N_0 B_d}{z_{2,2}(t)}.  \label{P_d_d}
\end{align}
Thus, the optimal transmit power vector is $\mathbf{P}_2 = [P_{b1, 2}^*, P_{b2,2}^*, P_{d,2}^*]$, where $P_{b2, 2}^* = 0$.

Similar as in cellular mode, there are 3 scenarios based on the values of $R_{d1}^*(t, \mathbf{P})$ and $R_{d2}^*(t, \mathbf{P})$:
\begin{enumerate}
\item If $R_{d1}^*(t, \mathbf{P}) \geq \alpha_1(t)$ and $R_{d2}^*(t, \mathbf{P}) \geq \alpha_2(t)$, let $pri(2) = 1$;

\item If $R_{d1}^*(t, \mathbf{P}) \geq \alpha_1(t)$ or $R_{d2}^*(t, \mathbf{P}) \geq \alpha_2(t)$, let $pri(2) = 2$;

\item If $R_{d1}^*(t, \mathbf{P}) < \alpha_1(t)$ and $R_{d2}^*(t, \mathbf{P}) < \alpha_2(t)$, let $pri(2) = 3$.
\end{enumerate}

\subsection{Reuse Mode}
Reuse mode is the most complicated case due to the impact of interference. From (\ref{R_r1}), (\ref{R_r2}) and (\ref{A_3}), the powers $P_{b1}(t) = P_1(t)$ and $P_{d}(t) = P_2(t)$ can be determined by having $R_{1,3}(t, \mathbf{P})$ and $R_{2,3}(t, \mathbf{P})$ attain their upper bounds $\beta_1(t)$ and $\beta_2(t)$ as follows:
\begin{align}
& B_r \log \left(1+\frac{P_1(t) z_{1,1}(t)}{P_2(t)z_{2,1}(t)+N_0B_r}\right) = \beta_1(t) \label{R_u11}\\
& B_r \log \left(1+\frac{P_2(t) z_{2,2}(t)}{P_1(t)z_{1,2}(t)+N_0B_r}\right) = \beta_2(t). \label{R_u21}
\end{align}
After simple algebraic steps, (\ref{R_u11}) and (\ref{R_u21}) can be rewritten as
\begin{align}
& P_1(t)z_{1,1}(t) - \big(2^{\frac{\beta_1(t)}{B_r}}-1\big)\big(P_2(t)z_{2,1}(t)+N_0B_r\big) = 0 \label{R_u12}\\
& P_2(t)z_{2,2}(t) - \big(2^{\frac{\beta_2(t)}{B_r}}-1\big)\big(P_1(t)z_{1,2}(t)+N_0B_r\big) = 0. \label{R_u22}
\end{align}
Since (\ref{R_u12}) and (\ref{R_u22}) constitute a system of linear equations with two unknowns, $P_1(t)$ and $P_2(t)$ can be derived in closed-form as follows:
\begin{align}
& P_1(t) = \frac{(2^{\frac{\beta_1(t)}{B_r}}-1)\big(z_{2,2}(t)+(2^{\frac{\beta_2(t)}{B_r}}-1)z_{2,1}(t)\big)N_0 B_r}
{z_{1,1}(t)z_{2,2}(t)-(2^{\frac{\beta_1(t)}{B_r}}-1)(2^{\frac{\beta_2(t)}{B_r}}-1)z_{1,2}(t)z_{2,1}(t)}               \\
& P_2(t) = \frac{(2^{\frac{\beta_2(t)}{B_r}}-1)\big(z_{1,1}(t)+(2^{\frac{\beta_1(t)}{B_r}}-1)z_{1,2}(t)\big)N_0 B_r}
{z_{1,1}(t)z_{2,2}(t)-(2^{\frac{\beta_1(t)}{B_r}}-1)(2^{\frac{\beta_2(t)}{B_r}}-1)z_{1,2}(t)z_{2,1}(t)}.
\end{align}

If $0 \leq P_1(t) \leq P_{bmax}$ and $0 \leq P_2(t) \leq P_{dmax}$, $\mathcal{P}^* = [P_1(t), 0, P_2(t)]$ is the optimal solution in reuse mode. Otherwise, under the constraints (\ref{P_b1}), (\ref{P_b2}), (\ref{P_d}) and (\ref{A_3}), the optimal solution $(P_{b1}^*, P_d^*)$ always satisfies either $P_{b1}^* = P_{bmax}$ or $P_{d}^* = P_{dmax}$. We can show that this leads to either $R_{1,3}(t, \mathbf{P}) < \beta_1(t)$ and $R_{2,3}(t, \mathbf{P}) < \beta_2(t)$ or $R_{1,3}(t, \mathbf{P}) = \beta_1(t)$ or $R_{2,3}(t, \mathbf{P}) < \beta_2(t)$. Hence, we need to consider both cases of $P_{b1}^* = P_{bmax}$ and $P_{d}^* = P_{dmax}$.

\subsubsection{Case 1: $P_{b1}^* = P_{bmax}$}
In this section, we analyze the strategy to find the optimal solution subject to $P_{b1}^* = P_{bmax}$. After setting $P_{b1}^* = P_{bmax}$, we can find the feasible region of $P_{d}$ by solving (\ref{A_3}) and (\ref{A_4}) as follows:
\begin{align}
& P_{dl1} \leq P_d \leq P_{dh1} \\
& P_{dl2} \leq P_d \leq P_{dh2}
\end{align}
where
\begin{align}
& P_{dl1} = \frac{P_{bmax} z_{1,1}(t)}{2^{\frac{\beta_1(t)}{B_r}} z_{2,1}(t)} - \frac{N_0 B_r}{z_{2,1}(t)} \\
& P_{dh1} = \frac{P_{bmax} z_{1,1}(t)}{2^{\frac{\max\{0, \alpha_1(t)\}}{B_r}} z_{2,1}(t)} - \frac{N_0 B_r}{z_{2,1}(t)} \\
& P_{dl2} = \frac{(2^{\frac{\max\{0, \alpha_2(t)\}}{B_r}}-1)(N_0 B_r + P_{bmax}z_{1,2}(t))}{z_{2,2}(t)} \\
& P_{dh2} = \frac{(2^{\frac{\beta_2(t)}{B_r}}-1)(N_0 B_r + P_{bmax}z_{1,2}(t))}{z_{2,2}(t)}.
\end{align}
Let
\begin{align}
& P_{dmin1} = \max\{0, P_{dl1}\} \\
& P_{dmax1} = \min\{P_{dmax}, P_{dh1}\} \\
& P_{dmin2} = \max\{0, P_{dl2}\} \\
& P_{dmax2} = \max\{P_{dmax}, P_{dh2}\} \\
& P_{dl} = \max\{P_{dmin1}, P_{dmin2}\} \\
& P_{dh} = \min\{P_{dmax1}, P_{dmax2}\}.
\end{align}

We again have 3 cases to consider:
\begin{enumerate}
\item If $P_{dl} \leq P_{dh}$, then the underflow and overflow playout buffer constraints are satisfied at both $C_1$ and $D_2$. This case has the highest priority, and let $pri_3(1) = 1$. We can show that the optimal solution always occurs at the endpoints. Therefore, the optimal transmit power vector is $\mathbf{P}_{3,1} = [P_{bmax}, 0, P_{dl}]$ if $R_{tot}(t, [P_{bmax}, 0, P_{dl}]) > R_{tot}(t, [P_{bmax}, 0, P_{dh})$; otherwise, $\mathbf{P}_{3,1} = [P_{bmax}, 0, P_{dh}]$.

\item If $P_{bmin1} \leq P_{bmax1}$ or $P_{bmin2} \leq P_{bmax2}$, then the underflow and overflow playout buffer constraints are satisfied at either only $C_1$ or $D_2$. For this case, we let $pri_3(1) = 2$. Similarly, there are four endpoint vectors $[P_{bmax}, 0, P_{d,j}]$ where $P_{d,1} = P_{dmin1}$, $P_{d,2} = P_{dmax1}$, $P_{d,3} = P_{dmin2}$ and $P_{d,4} = P_{dmax2}$. The optimal transmit power vector is $\mathbf{P}_{3,1} = [P_{bmax}, 0, P_{d,k}]$ if $R_{tot}(t, [P_{bmax}, 0, P_{d,k}])$ is the highest value among $R_{tot}(t, [P_{bmax}, 0, P_{d,j}])$ for all $j \in {1, 2, 3, 4}$.

\item If $P_{bmin1} > P_{bmax1}$ and $P_{bmin2} > P_{bmax2}$, then the underflow and overflow playout buffer constraints are not satisfied at $C_1$ and $D_2$. This case has the lowest priority, and we let $pri_3(1) = 3$. Similarly, there are two endpoint vectors, and the optimal transmit power vector is $\mathbf{P}_{3,1} = [P_{bmax}, 0, 0]$ if $R_{tot}(t, [P_{bmax}, 0, 0]) > R_{tot}(t, [P_{bmax}, 0, P_{dmax})$; otherwise, $\mathbf{P}_{3,1} = [P_{bmax}, 0, P_{dmax}]$.
\end{enumerate}

\subsubsection{Case 2: $P_{d}^* = P_{dmax}$}
Similarly as in the case of $P_{b}^* = P_{bmax}$, after fixing $P_{d}^* = P_{dmax}$, we can find the feasible region of $P_{b1}$ by solving (\ref{A_3}) and (\ref{A_4}):
\begin{align}
& P_{bl1} \leq P_{b1} \leq P_{bh1} \\
& P_{bl2} \leq P_{b1} \leq P_{bh2}
\end{align}
where
\begin{align}
& P_{bl1} = \frac{(2^{\frac{\max\{0, \alpha_1(t)\}}{B_r}}-1)(N_0 B_r + P_{dmax}z_{2,1}(t))}{z_{1,1}(t)} \\
& P_{bh1} = \frac{(2^{\frac{\beta_1(t)}{B_r}}-1)(N_0 B_r + P_{dmax}z_{2,1}(t))}{z_{1,1}(t)} \\
& P_{bl2} = \frac{P_{dmax} z_{2,2}(t)}{2^{\frac{\beta_2(t)}{B_r}} z_{1,2}(t)} - \frac{N_0 B_r}{z_{1,2}(t)} \\
& P_{bh2} = \frac{P_{dmax} z_{2,2}(t)}{2^{\frac{\max\{0, \alpha_2(t)\}}{B_r}} z_{1,2}(t)} - \frac{N_0 B_r}{z_{1,2}(t)}.
\end{align}
Let
\begin{align}
& P_{bmin1} = \max\{0, P_{bl1}\} \\
& P_{bmax1} = \min\{P_{bmax}, P_{bh1}\} \\
& P_{bmin2} = \max\{0, P_{bl2}\} \\
& P_{bmax2} = \max\{P_{bmax}, P_{bh2}\} \\
& P_{bl} = \max\{P_{bmin1}, P_{bmin2}\} \\
& P_{bh} = \min\{P_{bmax1}, P_{bmax2}\}.
\end{align}
There are also 3 cases:
\begin{enumerate}
\item If $P_{bl} \leq P_{bh}$, then the underflow and overflow playout buffer constraints are satisfied at both $C_1$ and $D_2$. This case has the highest priority, and let $pri_3(2) = 1$. We can show that the optimal solution always occurs at the endpoints. Therefore, the optimal transmit power vector is $\mathbf{P}_{3,2} = [P_{bl}, 0, P_{dmax}]$ if $R_{tot}(t, [P_{bl}, 0, P_{dmax}]) > R_{tot}(t, [P_{bh}, 0, P_{dmax})$; otherwise, $\mathbf{P}_{3,1} = [P_{bh}, 0, P_{dmax}]$.

\item If $P_{dmin1} \leq P_{dmax1}$ or $P_{dmin2} \leq P_{dmax2}$, then the underflow and overflow playout buffer constraints are satisfied at either only $C_1$ or $D_2$. For this case, we let $pri_3(2) = 2$. Similarly, there are four endpoint vectors $[P_{b1, j}, 0, P_{dmax}]$ where $P_{b1,1} = P_{bmin1}$, $P_{b1,2} = P_{bmax1}$, $P_{b1,3} = P_{bmin2}$ and $P_{b1,4} = P_{bmax2}$. The optimal transmit power vector is $\mathbf{P}_{3,2} = [P_{b1,k}, 0, P_{dmax}]$ if $R_{tot}(t, [P_{b1,k}, 0, P_{dmax}])$ is the highest value among $R_{tot}(t, [P_{b1,j}, 0, P_{dmax}])$ for all $j \in {1, 2, 3, 4}$.

\item If $P_{dmin1} > P_{dmax1}$ and $P_{dmin2} > P_{dmax2}$, then the underflow and overflow playout buffer constraints are not satisfied at $C_1$ and $D_2$. This case has the lowest priority, and we let $pri_3(2) = 3$. Similarly, there are two endpoint vectors, and the optimal transmit power vector is $\mathbf{P}_{3,2} = [0, 0, P_{dmax}]$ if $R_{tot}(t, [0, 0, P_{dmax}]) > R_{tot}(t, [P_{bmax}, 0, P_{dmax})$; otherwise, $\mathbf{P}_{3,2} = [P_{bmax}, 0, P_{dmax}]$.
\end{enumerate}

The overall optimal transmit power is selected from the above two cases in reuse mode. If $pri_3(i_1) < pri_3(i_2)$, the optimal transmit power vector is $\mathbf{P}_3 = \mathbf{P}_{3,i_1}$, where $i_1, i_2 \in {1, 2}$ and $i_1 \neq i_2$. Otherwise, $\mathbf{P}_3 = \mathbf{P}_{3,i_1}$ if $R_{tot}(t, \mathbf{P}_{3,i_1}) \geq R_{tot}(t, \mathbf{P}_{3,i_2})$. Let $pri(3) = pri_3(i_1)$ and the optimal transmit power vector is $\mathbf{P}_3 = \mathbf{P}_{3,i_1}$.

After determining the optimal transmit power vectors in three different transmission modes, we need to decide which mode to select as the best strategy for data transmission. Since we want to choose the one with the highest priority, if $pri(l) = \min\{pri(1), pri(2), pri(3)\}$ for only one value of $l \in {1,2,3}$, then the optimal transmit power vector is $\mathbf{P} = \mathbf{P}_l$ and the mode selection is $l$. And if $pri(1) = pri(2) = pri(3)$, then the one with the highest $R_{tot}(t, \mathbf{P}_{l})$ is chosen. Otherwise, if $pri(l_1) = pri(l_2) < pri(l_2)$, then $\mathbf{P} = \mathbf{P}_{l_1}$ if $R_{tot}(t, \mathbf{P}_{l_1}) \geq R_{tot}(t, \mathbf{P}_{l_2})$.

\section{Numerical and Simulation Results}\label{sec:Result}
In this section, we evaluate the performance of the proposed transmission strategies.
Rayleigh fading is considered in the channels in all simulations, where the normalized path gain is exponentially distributed as $f(z_{i,j}(t)) = \frac{1}{G_{i,j}}\exp\left\{\frac{-z_{i,j}(t)}{G_{i,j}}\right\}$ with path gain averages $G_{i,j}$, where $i \in \{1, 2\}$ and $j \in \{1, 2, 3\}$. The peak power constraints are $P_{dmax} = 0$ dB and $P_{bmax} = 2$ dB at $D_1$ and $BS$, respectively.
The movie \textit{Tokyo Olympics} is transmitted through cellular link and \textit{NBC News} is transmitted from $D_1$ to $D_2$ through different links according to the mode selection. The used VBR video traces in all the simulations are from the Video Trace Library hosted at Arizona State University \cite{reisslein}. The playout buffer size is set to be $1.5$ times the largest frame size in all the videos.

\begin{figure}
\centering
\includegraphics[width=0.4\textwidth]{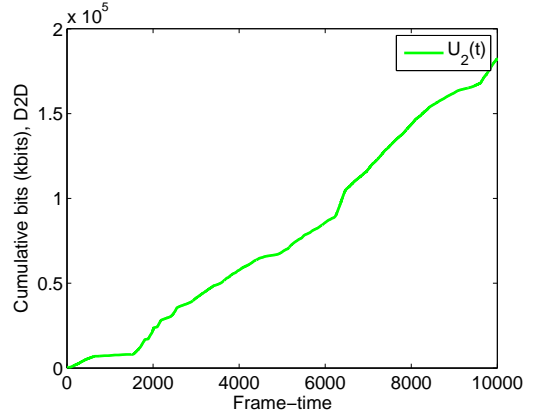}
\caption{\small{Consumption curve in $D_2$}}\label{fig:Cumu_DU}
\end{figure}

\begin{figure*}
\centering
\begin{subfigure}[b]{0.32\textwidth}
\centering
\includegraphics[width=\textwidth]{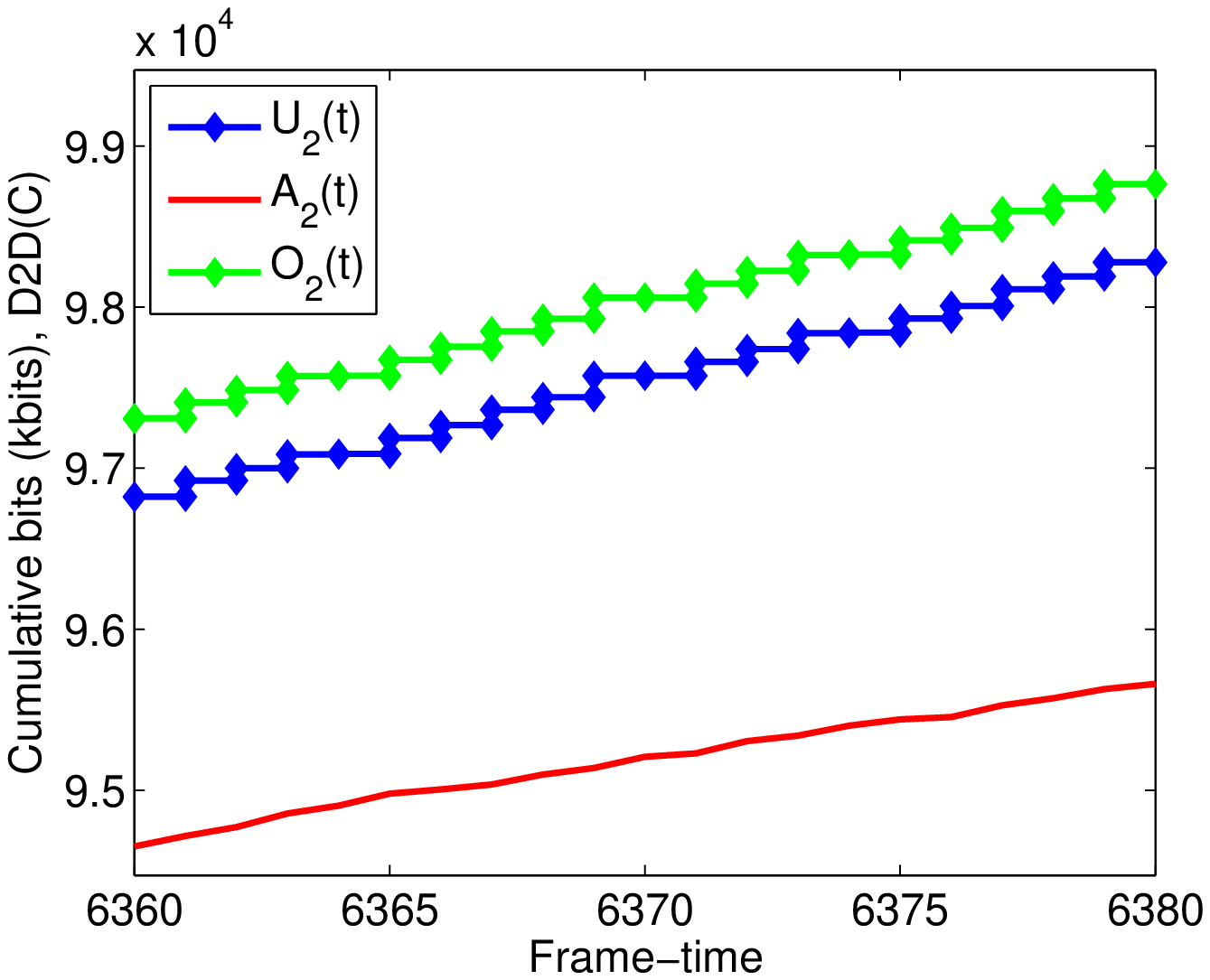}
\caption{Curves in cellular mode}\label{fig:f1:Cumu_Dc}
\end{subfigure}
\begin{subfigure}[b]{0.32\textwidth}
\centering
\includegraphics[width=\textwidth]{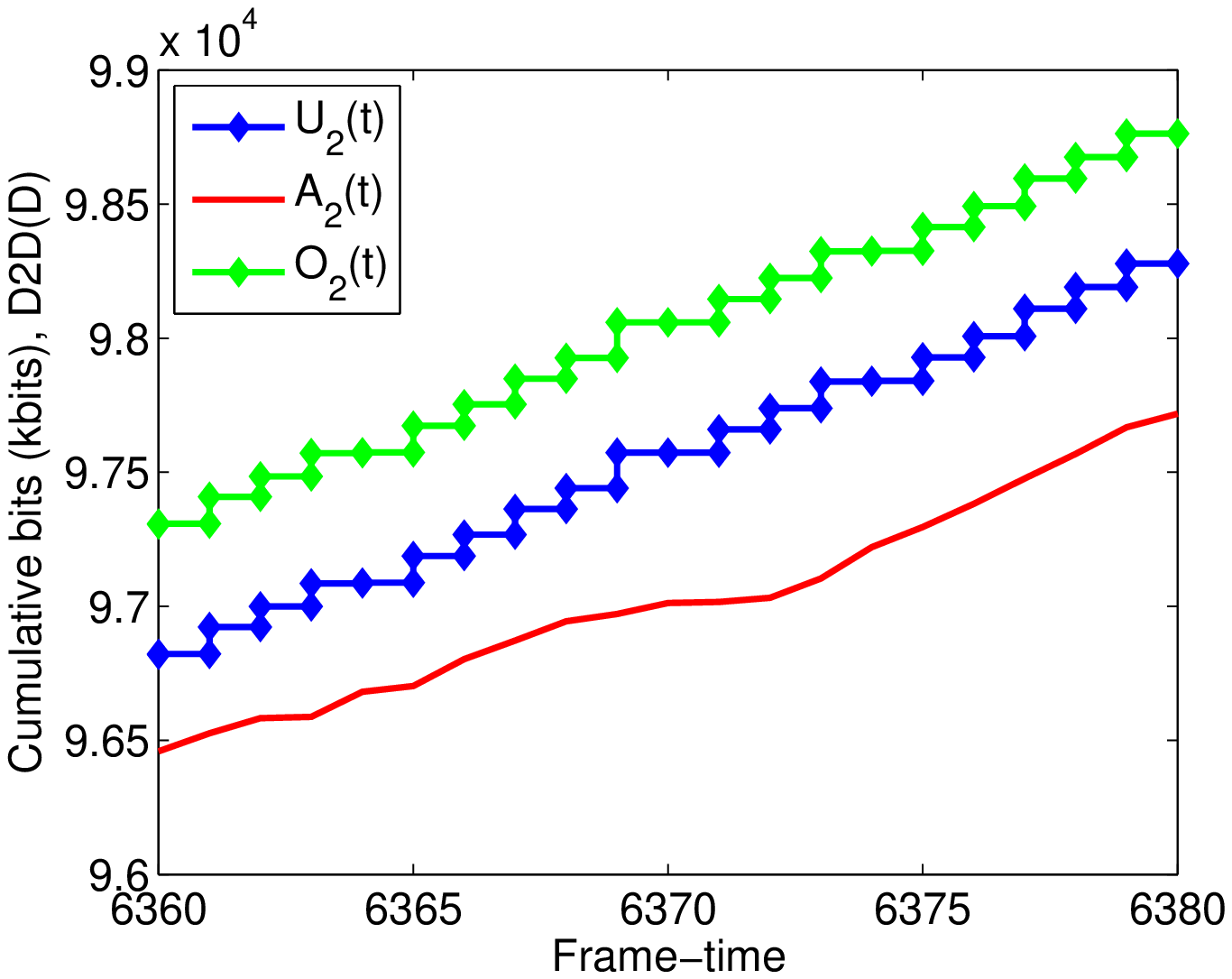}
\caption{Curves in dedicated mode}\label{fig:f1:Cumu_Dd}
\end{subfigure}
\begin{subfigure}[b]{0.32\textwidth}
\centering
\includegraphics[width=\textwidth]{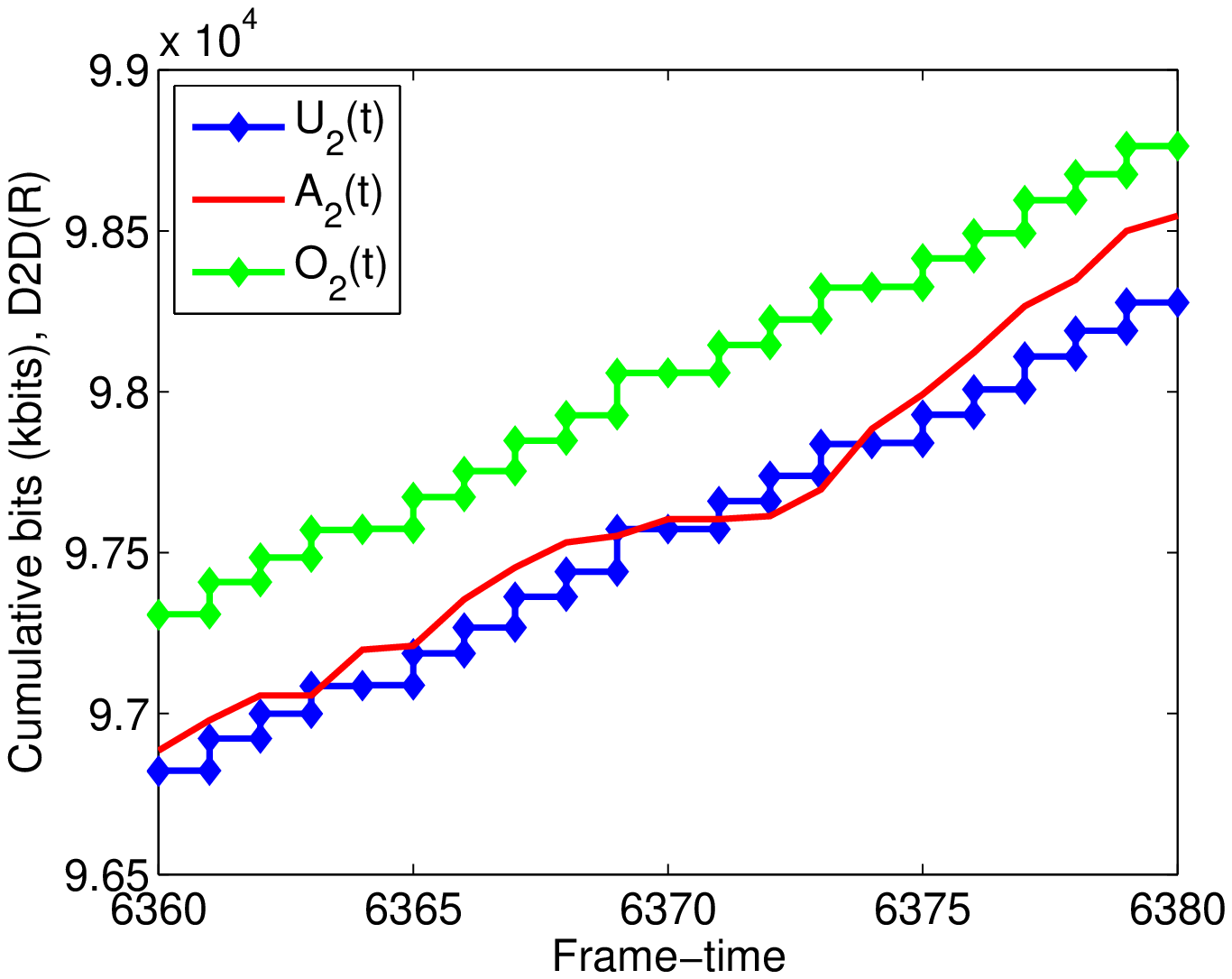}
\caption{Curves in reuse mode}\label{fig:f1:Cumu_Dr}
\end{subfigure}
\begin{subfigure}[b]{0.32\textwidth}
\centering
\includegraphics[width=\textwidth]{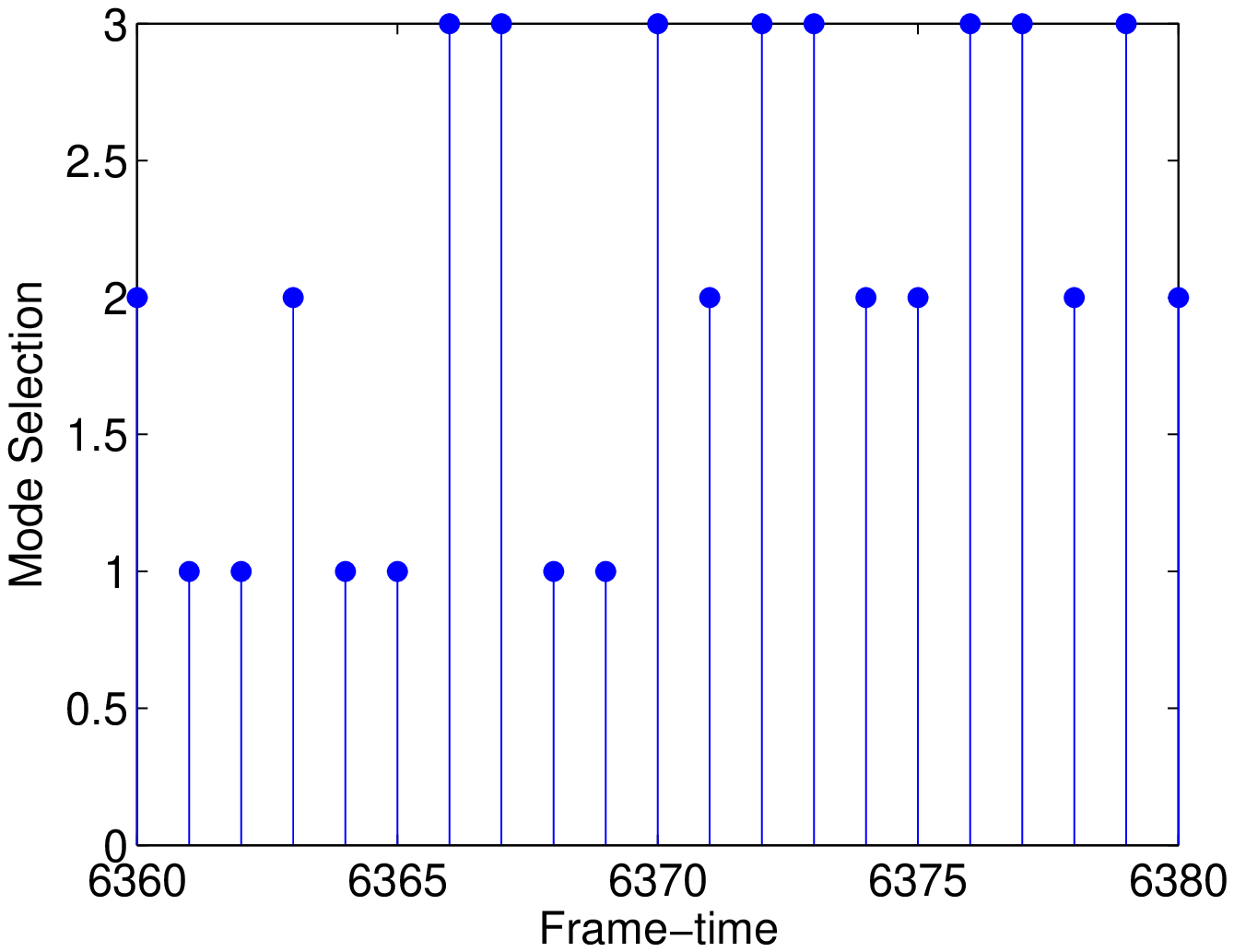}
\caption{Mode selection}\label{fig:f1:Mode}
\end{subfigure}
\begin{subfigure}[b]{0.32\textwidth}
\centering
\includegraphics[width=\textwidth]{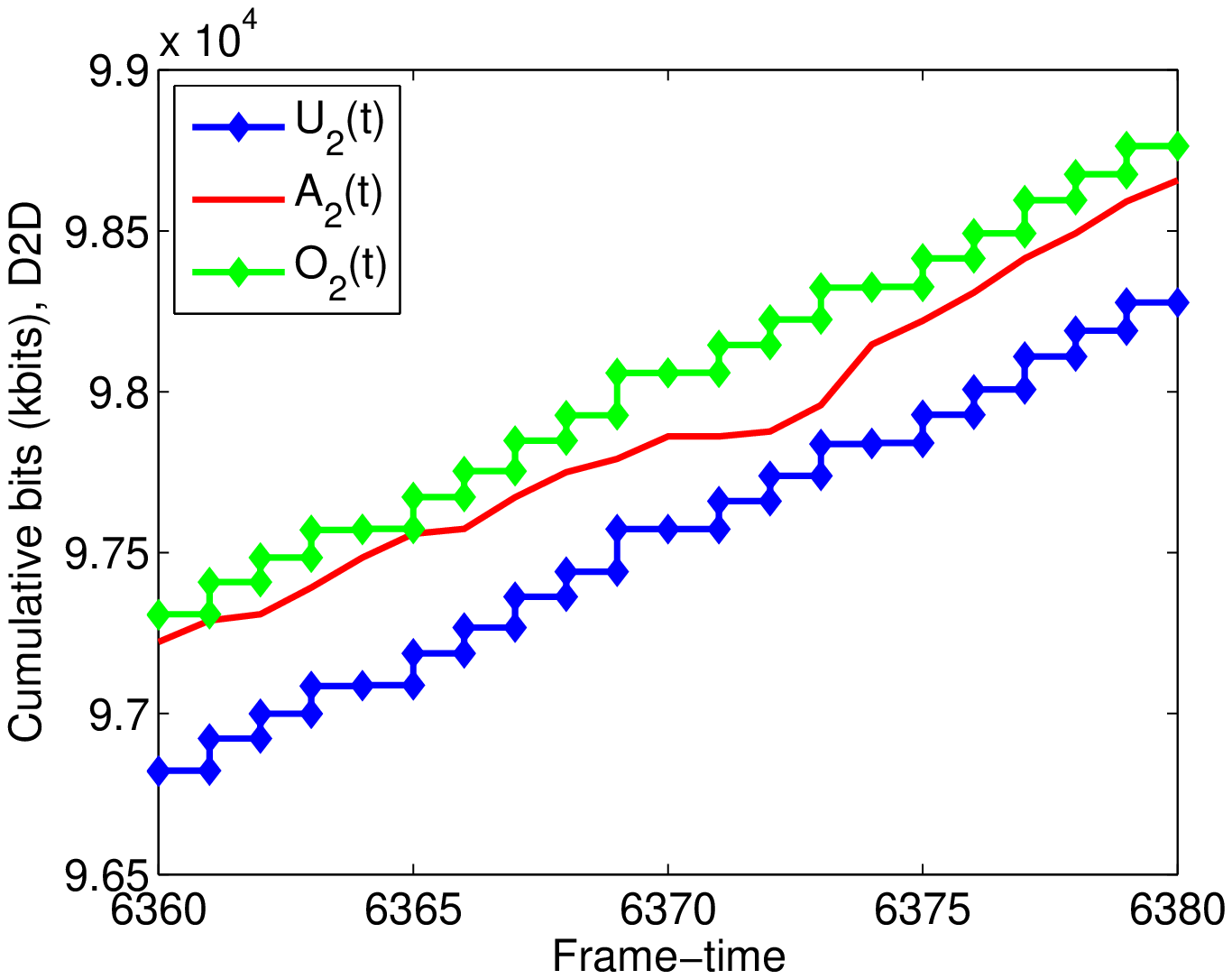}
\caption{Curves after mode selection}\label{fig:f1:Cumu_D}
\end{subfigure}
\caption{\small{The cumulative overflow, transmission, and consumption curves when transmitting \textit{NBC News} at D2D link in (a) cellular mode; (b) dedicated mode; (c) reuse mode and; (d) the optimal mode selection and (e) the corresponding curves with optimal mode selection.}}\label{fig:f1}
\end{figure*}

Fig. \ref{fig:Cumu_DU} shows the consumption curves of the buffer at $D_2$ from frame-time slot $1$ to $10000$. In Fig. \ref{fig:f1}, we plot the cumulative overflow, transmission, and consumption curves when transmitting  \textit{NBC News} between the D2D users from frame-time slot $6360$ to $6380$ in different transmission modes. In this time period, the cumulative transmission curves are lower than the cumulative consumption curves all the time in the cellular mode and dedicated mode. The reason is that all the curves are cumulative, and the cumulative transmission curves are much lower than the cumulative consumption curves before frame-time slot $6360$ and the transmit powers are not large enough for supporting the demanded data transmission. Additionally, the frame sizes around frame-time slot $6360$ are large. There are just several overflow events happening among these frame-time slots in the reuse mode since the cumulative transmission curve satisfies the buffer constraints before frame-time slot $6360$. The mode selection has the best performance since it always chooses the best transmission mode in each frame-time slot and this leads to the best cumulative transmission curve. The extra transmitted video data will be in the playout buffer to provide a cushion to variations in the network dynamics when future large frames need to be transmitted. From Fig. \ref{fig:f1:Cumu_D}, $A_2(t)$ at time slot $6360$ after mode selection is much higher than in cellular, dedicated and reuse modes. This advantage is due to the cumulative benefits. Fig. \ref{fig:f1:Mode} shows the mode selection after solving the optimization problem. $1$, $2$ and $3$ denote cellular mode, dedicated mode and reuse mode, respectively.

\begin{figure}
\centering
\includegraphics[width=0.4\textwidth]{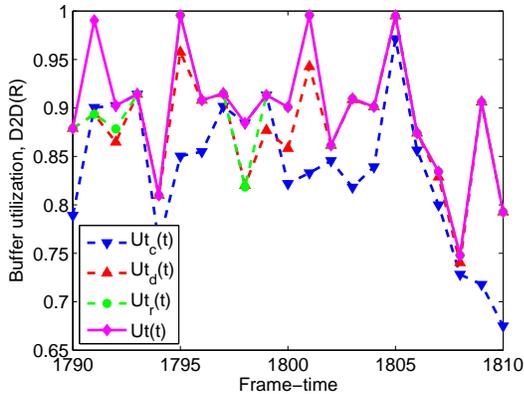}
\caption{\small{Buffer utilization in $D_2$}}\label{fig:Ut}
\end{figure}

Fig. \ref{fig:Ut} shows the buffer utilization from frame-time slot $1790$ to $1810$. We find that the buffer utilization of mode selection strategy has the highest value since this results as the solution of the optimization problem, and the values are higher than $70\%$. The higher buffer utilization leads to lower buffer underflow event probability. Table \ref{table_underflow} shows the probability of underflow events in different modes, and mode selection strategy has the lowest probability of underflow events both at $C_1$ and $D_2$.

\begin{table}[h]
\begin{center}
\caption{Probability of underflow events} \label{table_underflow}
    \begin{tabular}{| c | c | c | c | c |}
    \hline
         & cellular & dedicated & reuse & mode selection  \\ \hline
    $C_1$ & $0.0388$ & $0.0124$ & $3.6\times 10^{-4}$ & $3.6\times 10^{-4}$  \\ \hline
    $D_2$ & $0.0391$ & $0.0168$ & $0.0024$ & $1.6\times 10^{-4}$ \\ \hline
    \end{tabular}
\end{center}
\end{table}

\section{Conclusion}\label{sec:Conclusion}
In this paper, we have studied power control and mode selection for VBR video streaming in D2D networks. The problem formulation takes into account power control at the base station and device transmitter, the two-hop link model (as seen in cellular mode), interference (as experienced in reuse mode), VBR video characteristics and playout buffer requirements. We have proposed a low complexity strategy that can determine the optimal solution by comparing limited numbers of values out of which the best is chosen. The results demonstrate that the power control and mode selection strategy significantly improves the performance over just using a single mode. Specifically, power control and mode selection lead to better utilization of buffer and a smaller number of buffer overflows and underflows and video stall events, and hence provide improved quality of experience (QoE) to the users.


\begin{thebibliography}{1}
\bibitem{Cisco}
Cisco, ``Cisco visual networking index: Global mobile data traffic forecast update, 2015-2020 white paper,'' Feb 2016. [Online]. Available:
  \url{http://www.cisco.com/c/en/us/solutions/collateral/service-provider/visual-networking-index-vni/mobile-white-paper-c11-520862.html}


\bibitem{seetharma}
A.~Seetharam, P.~Dutta, V.~Arya, J.~Kurose, M.~Chetlur, and S.~Kalyanaraman,
  ``On managing quality of experience of multiple video streams in wireless
  networks,'' \emph{IEEE Transactions on Mobile Computing}, vol.~14, no.~3, pp.
  619--631, March 2015.

\bibitem{xu}
Y.~Xu, S.~E. Elayoubi, E.~Altman, R.~El-Azouzi, and Y.~Yu, ``Flow-level qoe of
  video streaming in wireless networks,'' \emph{IEEE Transactions on Mobile
  Computing}, vol.~15, no.~11, pp. 2762--2780, Nov 2016.

\bibitem{yhuang}
Y.~Huang and S.~Mao, ``Downlink power control for multi-user vbr video
  streaming in cellular networks,'' \emph{IEEE Transactions on Multimedia},
  vol.~15, no.~8, pp. 2137--2148, Dec 2013.

\bibitem{chatziperis}
S.~Chatziperis, P.~Koutsakis, and M.~Paterakis, ``A new call admission control
  mechanism for multimedia traffic over next-generation wireless cellular
  networks,'' \emph{IEEE Transactions on Mobile Computing}, vol.~7, no.~1, pp.
  95--112, Jan 2008.

\bibitem{rango}
F.~D. Rango, M.~Tropea, P.~Fazio, and S.~Marano, ``Call admission control for
  aggregate mpeg-2 traffic over multimedia geo-satellite networks,'' \emph{IEEE
  Transactions on Broadcasting}, vol.~54, no.~3, pp. 612--622, Sept 2008.

\bibitem{sen}
S.~Sen, D.~Towsley, Z.-L. Zhang, and J.~K. Dey, ``Optimal multicast smoothing
  of streaming video over the internet,'' \emph{IEEE Journal on Selected Areas
  in Communications}, vol.~20, no.~7, pp. 1345--1359, Sep 2002.

\bibitem{liang}
G.~Liang and B.~Liang, ``Balancing interruption frequency and buffering
  penalties in vbr video streaming,'' in \emph{IEEE INFOCOM 2007 - 26th IEEE
  International Conference on Computer Communications}, May 2007, pp.
  1406--1414.

\bibitem{reisslein}
M.~Reisslein, ``Video trace library.'' [Online]. Available:
  \url{http://trace.eas.asu.edu/}

\end{thebibliography}
\end{document}